\documentclass[genTeX]{nrc1}
\usepackage[pctex32]{graphicx}
\newcommand{\beq}{\begin{equation}}
\newcommand{\eeq}{\end{equation}}
\newcommand{\beqa}{\begin{eqnarray}}
\newcommand{\eeqa}{\end{eqnarray}}
\newcommand{\roughly}[1]{\mathrel{\raise.3ex\hbox{$#1$\kern-0.85em
\lower1ex\hbox{$\sim$}}}}
\newcommand{\gsim}{\roughly>}

\volyear{83}{2005}
\journal{Can. J. Phys.}
\received{June 5, 2005}
\accepted{?????}
\begin{document}

\title{Aspects of brane-antibrane inflation}
\author[James M.\ Cline]{James M.\ Cline}
\address{Department of Physics, McGill University, Montr\'eal, Qc 
H3A 2T8, Canada. \email{jcline@physics.mcgill.ca}}

\shortauthor{Cline}

\maketitle
\begin{abstract}
I describe a dynamical mechanism for solving the fine-tuning
problem of brane-antibrane inflation.  By inflating with stacks of
branes and antibranes, the branes can naturally be trapped at a
metastable minimum of the potential.  As branes tunnel out of this
minimum, the shape of the potential changes to make the minimum
shallower.  Eventually the minimum disappears and the remaining
branes roll slowly because the potential is nearly flat.  I show that
even with a small number of branes, there is a good chance of getting
enough inflation.  Running of the spectral index is correlated
with the tilt in such a way as to provide a test of the model by
future CMB experiments.
\end{abstract}

\section{Introduction}
\begin{figure}
\centerline{\includegraphics[width=0.5\hsize,angle=0]{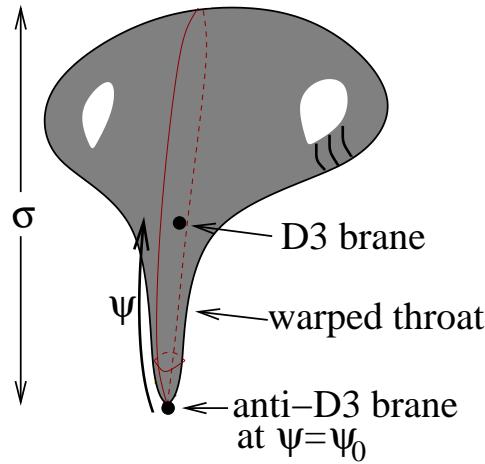}}  
\caption{The KKLMMT brane-antibrane inflation setup in a Calabi-Yau
space stabilized by fluxes and containing a warped throat.}
\label{fig0}
\end{figure}

Significant progress has been made recently in obtaining inflation
from specific string theory constructions, notably in type IIB
string theory in which moduli are stabilized by fluxes.  Aspects
of racetrack inflation, brane-antibrane inflation, and the problem
of reheating and cosmic string production in the latter, have been
reviewed elsewhere \cite{Cline}, and I will not repeat that material
here.  Instead I focus on a newer idea, multibrane inflation
\cite{CS}, which provides a simple and novel 
solution to the problem of fine-tuning of the inflaton potential in 
the KKLMMT \cite{KKLMMT} proposal for brane-antibrane inflation.
It is a straightforward generalization of the KKLMMT construction
shown in figure \ref{fig0}, where branes are attracted to antibranes
which live in the bottom of a Klebanov-Strassler \cite{KS} warped throat.

The Lagrangian for the inflaton field $\psi$, which is the 
brane-antibrane separation, is
\beq
\label{pot1}
	{\cal L} =  {1\over (2\sigma - \psi^2)^2}
\left(  6\sigma\dot\psi^2 - {\epsilon\, \tau\over  
1+ {\epsilon\over (\psi-\psi_0)^4}}\right)
\eeq
where $\epsilon$ is the fourth power of the warp factor at the
bottom of the Klebanov-Strassler throat, $\tau$ is the unwarped 3-brane
tension, $\sigma$ is the K\"ahler modulus of the Calabi-Yau manifold,
and $\psi_0$ is the position of the antibrane at the bottom of the
throat.  It was noted in \cite{BCSQ} that, depending on the values of
the parameters $\tau$, $\epsilon$, $\psi_0$, the inflaton may be
trapped in a metastable minimum near $\psi=0$.  For inflation,
one needs to tune the parameters to avoid this minimum, because
otherwise the inflaton does not roll, but rather undergoes old inflation with
its infamous graceful exit problem.  A tuning at the level of 1 part
in 1000 was needed to obtain sufficient inflation.

\begin{figure}
\centerline{\includegraphics[width=0.5\hsize,angle=0]{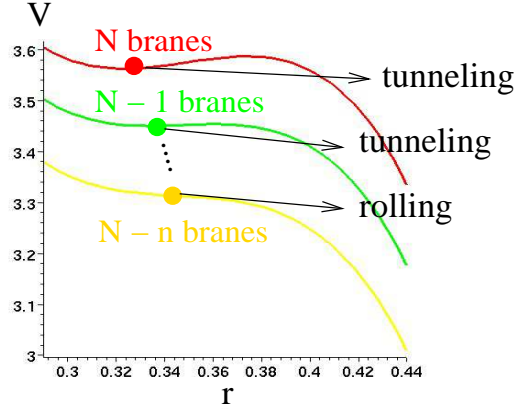}}  
\caption{Evolution of brane-antibrane potentials as number of trapped
branes decreases due to tunneling.}
\label{fig1}
\end{figure}

In this discussion, it was assumed that there is just a single brane
and antibrane driving inflation.  However in string theory it is
natural to consider stacks of coincident branes or antibranes.  If we
make this simple generalization, then a qualitatively new effect can
occur.  The potential (\ref{pot1}) generalizes to 
\beq
\label{potN}
	{\cal L} \to {1\over (2\sigma - \sum_i\psi_i^2)^2}
\left( \sum_i 6\sigma\dot\psi_i^2 - {N \epsilon\, \tau\over  
1+ \sum_i {\epsilon\over (\psi_i-\psi_0)^4}}\right)
\eeq
For some limited range of the parameter 
$b\equiv b\equiv {\sqrt{2\sigma\epsilon}\over \psi_0^3}$, this 
potential has the interesting property that for large a number $N$ of
branes, there is a metastable minimum near $\psi=0$, but the minimum
disappears as $N$ decreases.  This behavior is illustrated in figure
\ref{fig1}.  Therefore if one starts with a large enough number of branes,
then the successive tunneling of single branes from the metastable
minimum will cause the potential to automatically tune itself to
become increasingly flat, until the point at which the branes
can start to slowly roll and drive inflation.  During the tunneling
phase, ``old inflation'' is occurring, which by itself is not a 
successful inflationary scenario, but this is unimportant for us
since the last phase of old inflation is followed by inflation driven
by the slowly rolling branes. 

One way to understand the origin of this behavior
of the potential is to consider its curvature at the origin.
There are competing effects which depend
differently on $N$: one comes from the brane-antibrane Coulomb-like
potential, and the other comes from a supergravity effect, with the
entire flat-space potential being divided by the Calabi-Yau volume,
$2\sigma - N\psi^2$, as modified by the presence of the branes.
The curvature of the potential at $\psi=0$ is proportional to
\beq
V'' \sim {1\over {4\sigma}}-5/2\,{\frac {\epsilon}{{{ \psi_0}}^{2} \left( {{
 \psi_0}}^{4}+N\epsilon \right) }}+4\,{\frac {N{\epsilon}^{2}}{{{
\psi_0}}^{2} \left( {{ \psi_0}}^{4}+N\epsilon \right) ^{2}}}
\eeq
Clearly this is always positive for large enough $N$.  But when $N=0$,
it is proportional to 
\beq
V'' \sim {1\over {4\sigma}}-5/2\,{\frac {\epsilon}{{{ \psi_0}}^{6}}}
\eeq
which will be negative if $b^2\equiv 2\sigma\epsilon/\psi_0^6 > 0.2$ 
Thus the curvature of the potential
changes sign as a function of $N$. This
turns out to be just one condition on $b$ which must be satisfied in
order to get the right behavior.  
We will show how to get around 
this restriction later.	

We have explored the parameter space of $\sigma$ and $\epsilon$
(the size of the internal manifold and the warping in the inflationary
throat) to determine how many branes are needed, and how much
inflation results.  The correlation between the number of $e$-foldings
and the number of branes is shown in figure \label{fig2},  which was obtained
by uniformly sampling the parameter space $-5 < \log_10\epsilon < -2$
and $1 < \log_10(2\sigma) < 4$.  This corresponds to a moderately
warped inflationary throat, and large Calabi-Yau volumes ($2\sigma$ is
the size of the manifold in string units), which is required for
consistency of the low-energy supergravity description.  The
figure shows the actual points as well as an empirical formula which
explains the complicated structure rather well.  The origin of the
steep lines of points is due to the fact that, if $N$ could be
noninteger, it would be possible to potentials with perfectly flat
spots (where $V'$ and $V''$ vanish simultaneously) leading to very
long periods of inflations.  The lines are sequences of points in
which the actual value of $N$ becomes increasingly close to the 
ideal value.

\begin{figure}
\centerline{\includegraphics[width=0.7\hsize,angle=0]{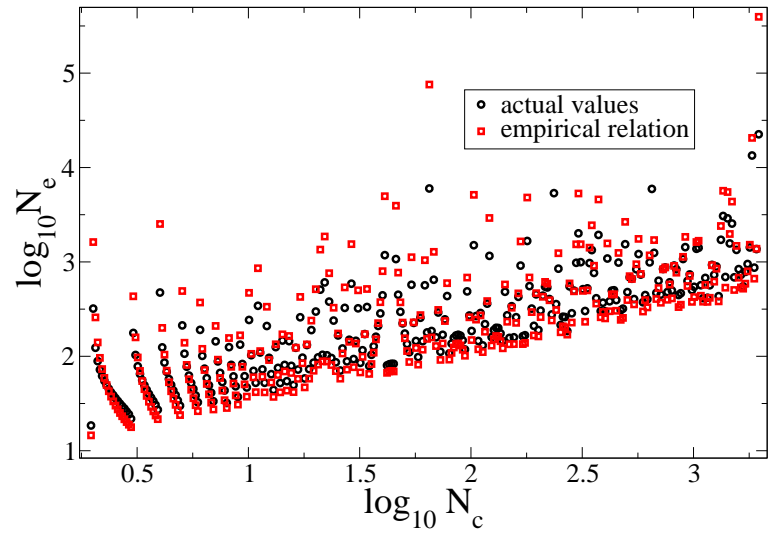}}  
\caption{Number of e-foldings of inflation versus critical number
of branes needed for a flat potential, from scanning over the
$\sigma$-$\epsilon$ parameter space.}
\label{fig2}
\end{figure}

\begin{figure}
\centerline{\includegraphics[width=0.7\hsize,angle=0]{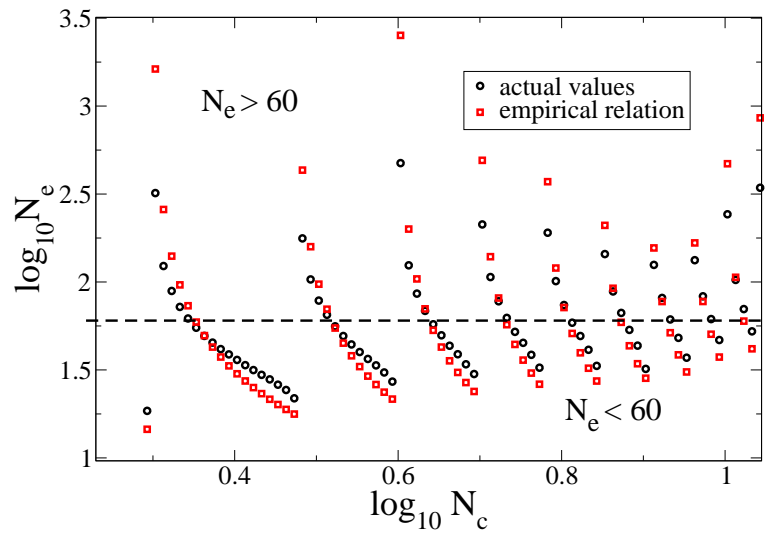}}  
\caption{Close-up of small-$N$ region of figure \ref{fig2}.}
\label{fig3}
\end{figure}

It is interesting to note that, even though we consider the
possibility of large numbers of branes, there is a high probability of
getting enough inflation $N_e \gsim 60$ even with a relatively small
number of branes, of order 10.  This is shown in figure \ref{fig3},
which is a close-up of the small-$N$ region of figure \ref{fig2}.
Nevertheless, brane stacks of up to 100 or even 1000 could be used
to satisfy the tadpole conditions of the flux compactification
scenario, given a Calabi-Yau manifold with large enough Euler
characteristic \cite{Shamit}.  Examples with $\chi = 10^5$ are known.

I come back to the fine-tuning issue.  If we had no restriction
on the combination $b\equiv\sqrt{2\sigma\epsilon}/\psi_0^3$, this
mechanism would be a complete solution to the problem of fine-tuning.
The fact that $b$ must lie in a narrow range makes it less compelling.
However, we have found that by adding corrections to the potential 
which are generically expected to be present, the restriction on 
$b$ can be removed altogether.  These take the form of corrections
to the superpotential and K\"ahler potential.   The potential 
generalizes to 
\beq
	V \to 
{N\delta(\psi^2-\psi_0^2)^2  - {N \epsilon\, \tau\over  
1+ N{\epsilon (\psi-\psi_0)^{-4}}} \over 
(2\sigma - N(\psi^2+d\psi^4))^2}
\eeq
Then using, for example,
$d = -0.5\psi_0^{-2}$ and $\delta = 0.1\epsilon\tau\psi_0^{-4}$,
we get the required self-flattening behavior of the potential for
any value of $b$.

The question of whether this theory of inflation can be distinguished
from others by the data is difficult.  In our scan of parameter
space, we find a wide range of predictions for properties of the
scalar and tensor perturbation spectra relevant for the Cosmic
Microwave Background.  For example, applying the COBE normalization
to the spectrum implies that a range of energy scales
for the potential is allowed, $V^{1/4} < 10^{-3} M_p$, with the
inequality being saturated only for the rare cases with $N\sim 1$
branes.  For this reason, a measurable tensor component in the CMB is
not a generic prediction of the model.  Moreover we find a wide range
of spectral tilts,  $0.93 < n_s < 1.15$, although the majority of
realizations has $n_s < 1$.  Thus there is no smoking-gun prediction
for the spectral index in this model.  

\begin{figure}
\centerline{\includegraphics[width=0.7\hsize,angle=0]{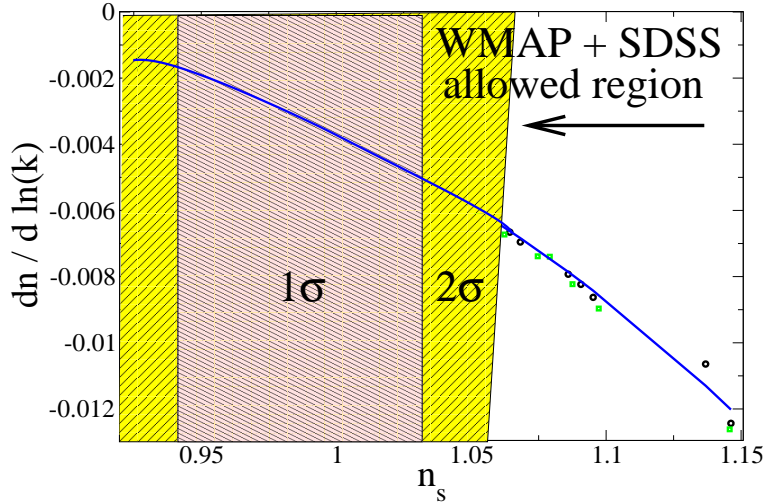}}  
\caption{Running of spectral index versus spectral index in multibrane
model.}
\label{fig4}
\end{figure}

However, there is an interesting test which could allow future
CMB experiments to rule out the models under consideration.  There
is presently a hint in the WMAP data of strong running of the spectral
index, $dn_s/d\ln k \sim -0.1$.  Our models were not able to produce
such a large amount of running, and the exact value awaits
experimental confirmation because the errors are still large.  But
if upcoming measurements were able to confirm even a much smaller
level of running,  $dn_s/d\ln k \sim -0.01$, it would effectively 
rule out the model because of the very definite correlation between
$dn_s/d\ln k$ and  the spectral index itself, $n_s$, as shown in
figure \ref{fig4}.  The predicted points fall along the solid line.
Although many of these points lie within the experimentally allowed
region \cite{Seljak}, those with significant running do not.  

In summary, we have presented a novel mechanism for solving the
fine-tuning problem of obtaining a flat inflaton potential, within
a popular string theoretic model of inflation.  The self-flattening
feature of the brane-antibrane potential does not naturally occur in
field theory models of inflation, but it appears quite generically
in the present stringy context.  It is also encouraging that at least
one experimental test can potentially falsify the model.


\begin{thebibliography}{9}
\bibitem{Cline}
  J.~M.~Cline,
  ``Inflation from string theory,''
  arXiv:hep-th/0501179.
  %%CITATION = HEP-TH 0501179;%%


\bibitem{CS}
  J.~M.~Cline and H.~Stoica,
  ``Multibrane inflation and dynamical flattening of the inflaton potential,''
  arXiv:hep-th/0508029.
  %%CITATION = HEP-TH 0508029;%%

\bibitem{KKLMMT}
  S.~Kachru, R.~Kallosh, A.~Linde and S.~P.~Trivedi,
  ``De Sitter vacua in string theory,''
  Phys.\ Rev.\ D {\bf 68}, 046005 (2003)
  [arXiv:hep-th/0301240].
  %%CITATION = HEP-TH 0301240;%%

\bibitem{KS}
  I.~R.~Klebanov and M.~J.~Strassler,
  ``Supergravity and a confining gauge theory: Duality cascades and
  chiSB-resolution of naked singularities,''
  JHEP {\bf 0008}, 052 (2000)
  [arXiv:hep-th/0007191].
  %%CITATION = HEP-TH 0007191;%%



\bibitem{BCSQ}
  C.~P.~Burgess, J.~M.~Cline, H.~Stoica and F.~Quevedo,
  ``Inflation in realistic D-brane models,''
  JHEP {\bf 0409}, 033 (2004)
  [arXiv:hep-th/0403119].
  %%CITATION = HEP-TH 0403119;%%

\bibitem{Shamit}
S.\ Kacrhu, private communication

\bibitem{Seljak}
  U.~Seljak {\it et al.},
  ``Cosmological parameter analysis including SDSS Ly-alpha forest and  galaxy
  bias: Constraints on the primordial spectrum of fluctuations,  neutrino mass,
  and dark energy,''
  Phys.\ Rev.\ D {\bf 71}, 103515 (2005)
  [arXiv:astro-ph/0407372].
  %%CITATION = ASTRO-PH 0407372;%%


\end{thebibliography}
\end{document}